# Hydrogen emission under laser exposure of colloidal solutions of nanoparticles

E.V.Barmina[1], A.V. Simakin[1], G.A. Shafeev[1,2]

[1]Wave Research Center of A.M. Prokhorov General Physics Institute of the Russian Academy of Sciences, 38, Vavilov street, 119991 Moscow Russian Federation

[2]National Research Nuclear University MEPhI (Moscow Engineering Physics Institute), 31, Kashirskoye highway, 115409, Moscow, Russian Federation

Abstract

We report the generation of molecular hydrogen from water by laser irradiation, without any electrodes and photocatalysts. A near infrared pulsed nanosecond laser is used for exposure of colloidal solution of Au nanoparticles suspended in water. Laser exposure of the colloidal solution results in formation of plasma of laser breakdown of liquid and emission of $H_2$. The rate of $H_2$ emission depends critically on the energy of laser pulses. There is a certain threshold in laser fluence in liquid (around 50 J/cm$^2$) below which plasma disappears and $H_2$ emission stops. $H_2$ emission from colloidal solution of Au nanoparticles in ethanol is higher than that from similar water colloid. It is found that formation of plasma and emission of $H_2$ or $D_2$ can be induced by laser exposure of pure liquids, either $H_2O$ or $D_2O$, respectively. The results are interpreted as water molecules splitting by direct electron impact from breakdown plasma.

## Introduction

Laser ablation of solids in liquids is a physical method of generation of large variety of nanoparticles. Typically, a solid target is placed into liquid, which is transparent for laser radiation. If the laser fluence on the target is enough for its melting, then this melt is dispersed into surrounding liquid as nanoparticles (NPs). Virtually any liquid is suitable for generation of NPs in this way, the most common of them is water. Organic solvents, such as alcohols, are also frequently used for preparation of colloidal solutions of different NPs in them. Laser ablation of solids in liquids is accompanied by formation of plasma plume above the target. Plasma is also observed under laser exposure of colloidal solutions of NPs. If laser intensity is high enough for NPs to reach temperatures of about $10^4$-$10^5$ K, some part of their atoms may be ionized [1]. When the fraction of these atoms reaches the critical value, plasma formation occurs [2]. Such type of plasma is often referred to as "nanoplasma" [3] due to its confinement in the small region in the NP vicinity. At high laser pulse repetition rates these small sources of plasma may unite between each other resulting in the liquid breakdown plasma formation. This means that the medium inside the laser beam undergoes strong overheating and ionization. In these conditions both the liquid and the material of NPs are affected resulting in chemical changes of their composition. In case of organic solvents, e.g., ethanol, this leads to deep pyrolysis of the liquid down to formation of elementary glassy carbon [4]. In [5] the formation of hydrogen gas from a mixture of pure carbon powder and water via laser irradiation was reported with intense nanosecond laser pulses without any electrodes or photocatalysts. The authors suggest that carbon acts like a catalyst, since gaseous products contain $CH_4$, CO, $CO_2$, and $H_2$. The peak power density of laser radiation is of order of 100 mJ/cm$^2$, which is apparently insufficient for water breakdown. However, presence of carbon in reaction products indicates that carbon is not just a catalyst but rather initial component of the reaction.

The aim of the present work is search of one of most probable product of solvents degradation, namely, hydrogen. For this purpose we exposed colloidal solutions of Au NPs in water and ethanol to pulsed radiation of a Nd:YAG laser. Presence of NPs provokes laser breakdown of the liquid and formation of microscopic plasma channel in it.

Experimental technique

Colloidal solutions of Au NPs were prepared using the technique of laser ablation in liquid. For this purpose, an Ytterbium fiber laser with pulse duration of 70 ns, repetition rate of 20 kHz and pulse energy of 1 mJ at 1060-1070 nm was used as irradiation source for NPs generation. Laser radiation was focused on metallic plate made of corresponding metal immersed into deionized water by an F-Theta objective. Laser beam was scanned across the sample surface at the speed of 1000 mm/s by means of galvo mirror system. Size distribution of obtained NPs was analyzed with Measuring Disc Centrifuge (CPS Instruments). Initial colloidal solution of Au NPs typically contained about $10^{12}$ NPs/ml with maximum of their size near 15 nm.  Further irradiation of NPs colloidal solutions in absence of the target and diluted in necessary proportion was carried out using the radiation of the Nd:YAG laser at wavelength of 1064 nm and pulse duration of about 10 ns (FWHM).  Laser radiation was focused inside the liquid by an F-Theta objective with focal distance of 90 mm (Fig. 1). Laser beam was scanned across the window along circular trajectory about 20 mm in diameter at the velocity of 1000 – 3000 mm/s by means of galvo mirror system. Laser exposure of 2 ml portions of colloids was carried out at 2 mJ energy per pulse and repetition rate of laser pulses of 10 kHz. Estimated diameter of the laser beam waist was 50 μm, which corresponds to laser fluence in the liquid of 80 J/cm$^2$. Bright cylinder made of plasma appeared 2-3 mm above the window inner surface and looks continuous for eye.

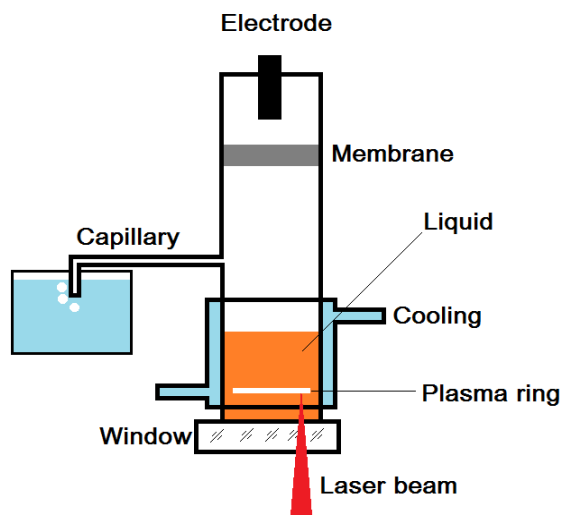

Fig. 1. Experimental setup for hydrogen emission.

Amperometric hydrogen sensor was used to monitor the concentration of $H_2$ in the space above the liquid surface. Excessive pressure in the system was released to ambient air through a glass capillary dipped in 1 cm thick water layer. In this case the total pressure in the cell was equal to atmospheric one. Inner electrolyte of the sensor is separated from the cell atmosphere by a membrane pervious only to $H_2$. The sensor indicates either the concentration of $H_2$ in µg/l or its partial pressure in Torrs. Calibration of the sensor was performed in air (no $H_2$) and in 1 atmosphere of $H_2$. The precision of $H_2$ concentration measurements is 5%. Total volume of the atmosphere above the water level can be estimated as 10 ml. The characteristic time of sensor response in current geometry is about 5 min.

Results

Laser exposure of the colloidal solution is accompanied by formation of plasma. Its view is shown in Fig. 2.

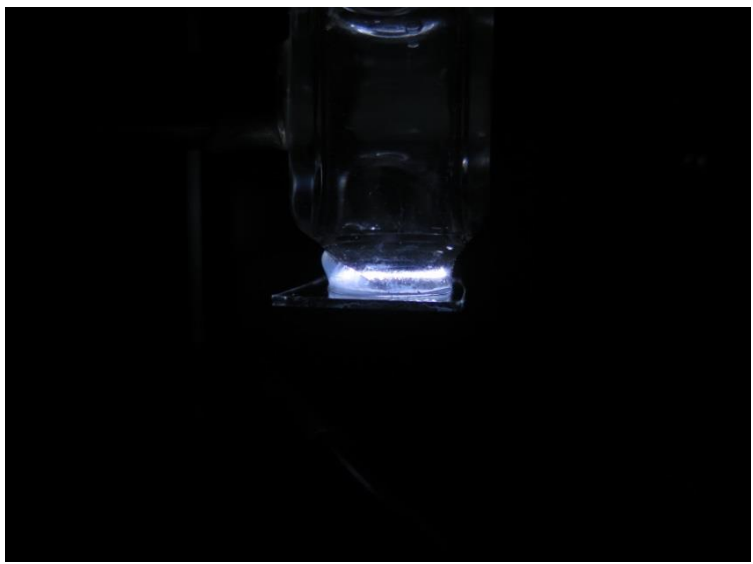

Fig. 2. Emission of plasma ring during laser exposure of the solution. Repetition rate of 10 kHz, scanning velocity of 1000 mm/s, camera exposure of 1/13 s, ring diameter of 20 mm.

Typical evolution curve of $H_2$ concentration under exposure of the colloidal solution of Au NPs is presented in Fig. 3, which shows three consecutive laser exposures of the same colloidal solution. As one can see, the emission of $H_2$ is well-reproducible. Hydrogen detector was purged by air between exposures. Typically, partial pressure of $H_2$ amounts to $200 - 300$ Torr. For the given setup the highest rate of $H_2$ emission is achieved at 10 min of laser exposure. Au NPs in the laser-exposed solution are fragmented to smaller NPs [6] with measured final size of few nm.

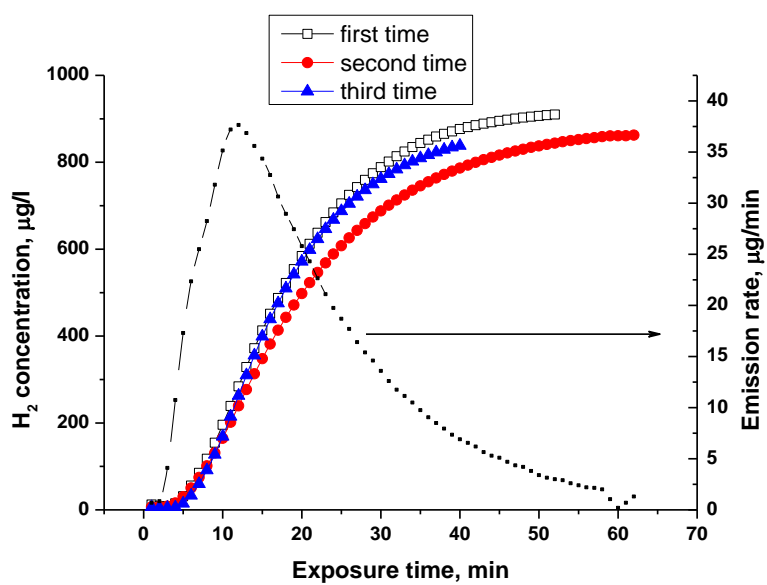

Fig. 3. Time dependence of $H_2$ concentration under 3 consecutive laser exposures of the same colloidal solution. Scanning velocity of 1000 mm/s, energy per pulse is 2 mJ. Dashed line is the rate of hydrogen emission for the second laser exposure.

It was found that the stationary concentration of emitted $H_2$ depends on the concentration of Au NPs at otherwise equal experimental conditions. Namely, the concentration of $H_2$ decreases with the increase of concentration of Au NPs (Fig. 4).

Laser exposure of the liquid is accompanied by formation multiple gas bubbles that ascends to the liquid surface. From time to time during exposure these bubbles can enter to the laser beam. This leads to micro-explosion of bubbles with well-discernable sound.

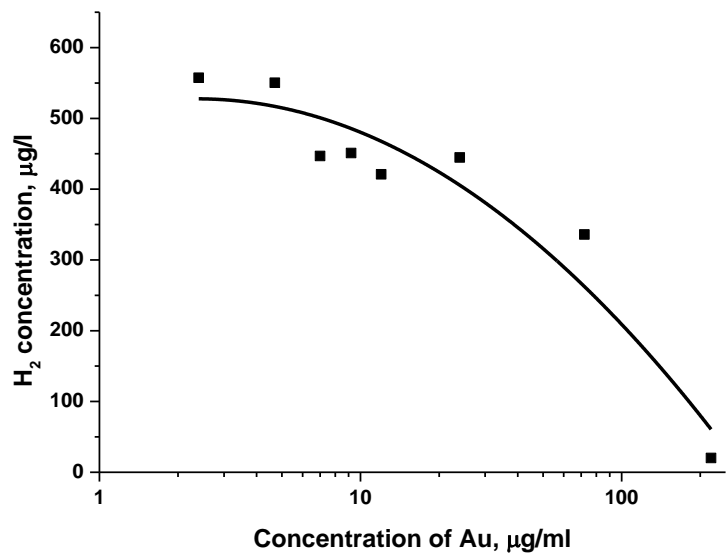

Fig. 4. Dependence of $H_2$ concentration after 30 minutes of laser exposure on the concentration of Au NPs in water.

The decrease of $H_2$ content with the increase of concentration of NPs is due to both scattering and absorption of laser radiation in water by NPs. As a result, the peak intensity of laser radiation inside the beam waist decreases, and plasma formation is dumped. Similar tendency is shown in Fig. 5 for various levels of laser pulse energy with the same position of the laser beam waist with respect to window.

Interestingly, $H_2$ emission can also be observed under laser exposure of technical water that contains some solid impurities, of order several mg per liter (Fig. 5).

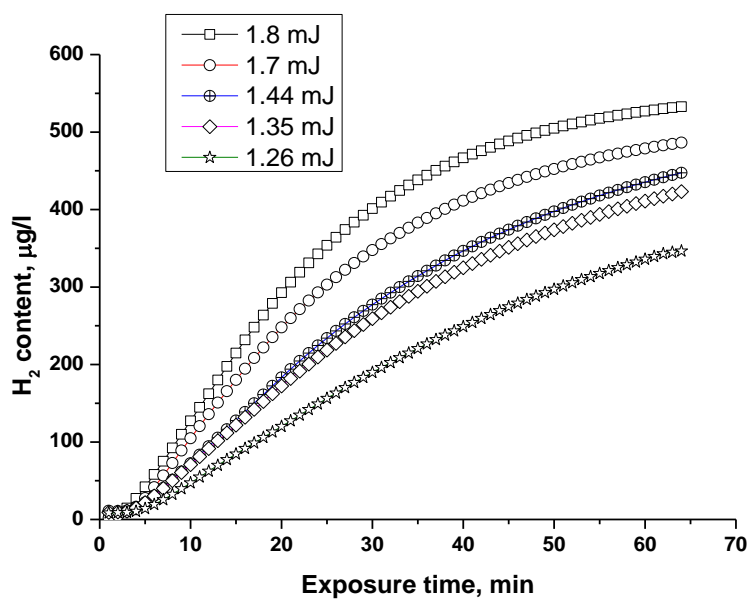

Fig. 5. Dependence of $H_2$ content above the liquid on exposure time of laser irradiation at various values of energy per pulse. Scanning velocity of the beam of 3000 mm/s, technical water.

One can see that there is laser power threshold for hydrogen emission. Below certain threshold the breakdown of technical water disappears, and emission of $H_2$ stops (Fig. 6).

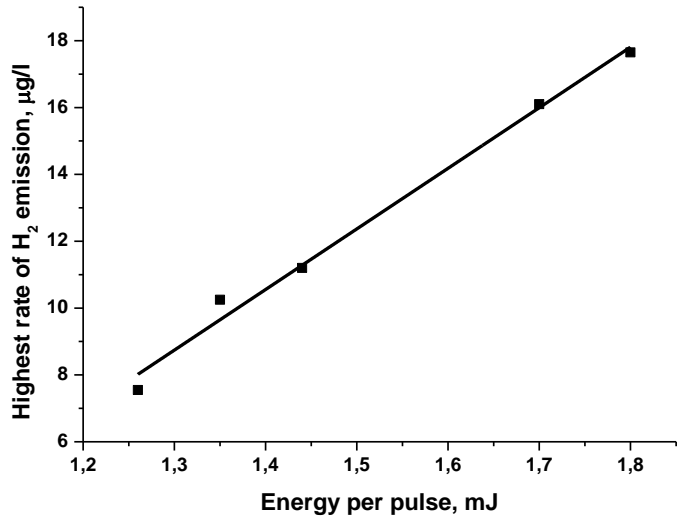

Fig. 6. The maximal rate of $H_2$ emission on the laser energy per pulse. Technical water, scanning velocity of 3000 mm/s.

On the other hand, the average rate of $H_2$ production over 1 hour laser exposure shows somewhat different behavior (see Fig. 7).

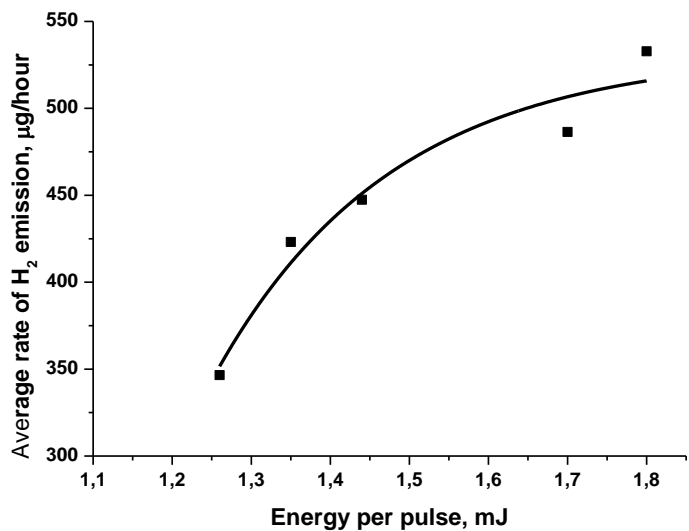

Fig. 7. Average hydrogen emission rate as a function of laser energy per pulse. Technical water, scanning velocity of 3000 mm/s.

The rate of hydrogen emission in colloidal solution of Au NPs in ethanol is much higher than in water. This is clearly seen at exposure times corresponding the maximum rate of $H_2$ emission (Fig. 8). Probably, this is due to pyrolysis of this solvent. Of course, the concentration of $H_2$ depends on the concentration of Au NP for both solvents. This dependence for water is shown in Fig. .

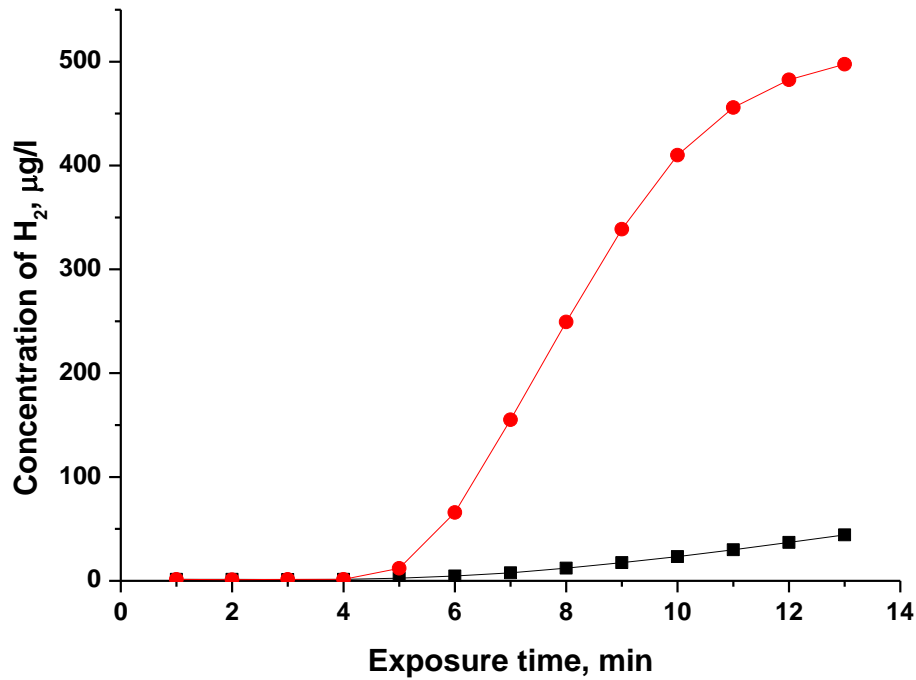

Fig. 8. Time dependence of $H_2$ concentration in colloidal solution of Au NPs in water (squares) and absolute ethanol (circles). Scanning velocity of 3000 mm/s, concentration of Au NPs in water of $10^{10}$ cm$^{-3}$.

Finally, it was found that the addition of Au NPs to water is not necessary for $H_2$ production. The dependence of $H_2$ emission rate on the relative concentration of Au NPs in water is presented in Fig. 9. Correspondingly, the stationary levels of $H_2$ concentration are achieved in longer time and are lower than in optimal case with high emission rate. One can see that the rate of $H_2$ production is not zero even without intentionally added Au NPs.

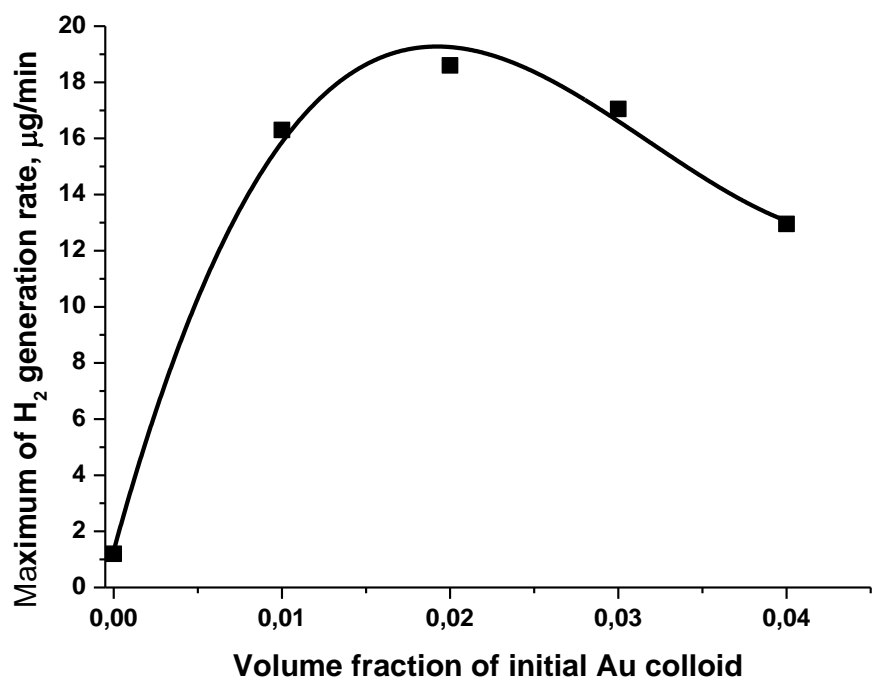

Fig. 9. Maximal rate of $H_2$ emission as the function of relative fraction of Au NPs. Zero fraction corresponds to water purified with reverse osmosis.

Similar result was obtained under laser exposure of saline solution in its original glass ampoule (Fig. 10).

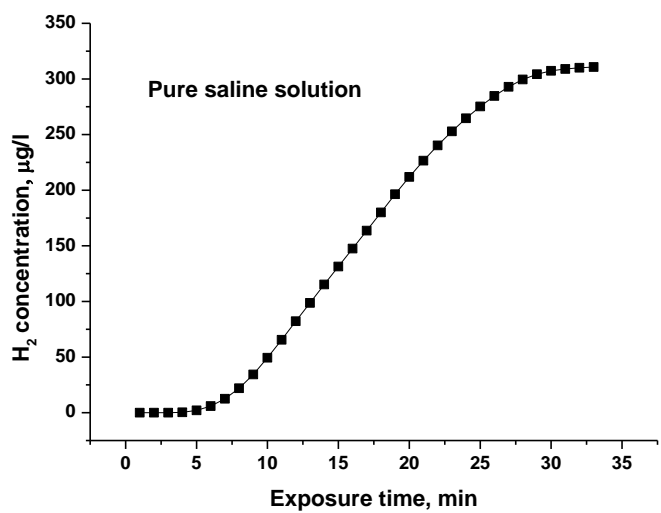

Fig. 10. Dynamics of $H_2$ emission during laser exposure of pure saline solution (0.7 % NaCl in water). 2 mJ per pulse.

H$_2$ emission is also observed under laser exposure of high purity water with conductivity of 0.8 μSm, in which the presence of solid impurities is vigorously controlled and is not possible, since this water is used in pharmaceutical industry.

Amperometric hydrogen sensor is not sensitive to isotopes of hydrogen, therefore, in case of laser exposure of heavy water D$_2$O it should detect gaseous deuterium. This indeed occurs. Fig. 11 shows the evolution of D$_2$ emission in case of laser exposure of pure D$_2$O.

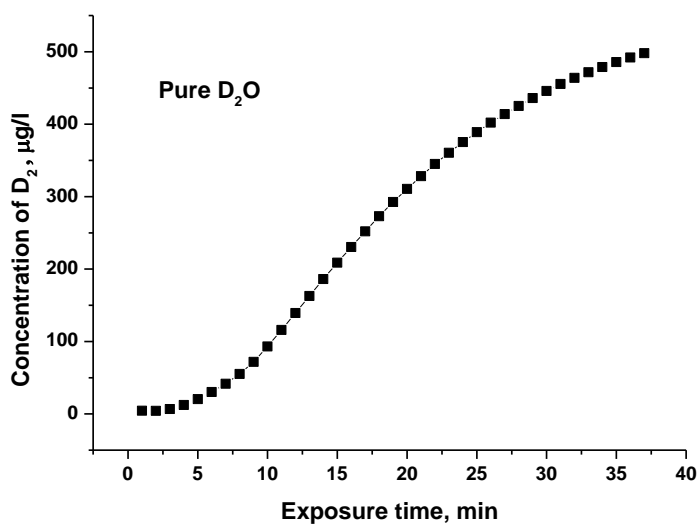

Fig. 11. Evolution of D$_2$ concentration under laser exposure of pure D$_2$O. The stationary level corresponds approximately to 250 Torr of D$_2$ partial pressure in the system.

Discussion

There are several possible mechanisms of H$_2$ emission in the present experimental conditions. Multi-photon absorption of water molecules is hardly possible due to small energy of photons (1 eV) and relatively long laser pulses. Estimated level of peak intensity in the liquid is $8 \times 10^9$ W/cm$^2$, which is quite moderate. Au NPs are not photocatalytically active. The most probable mechanism of water dissociation in our case is splitting of water molecules by electron impact. Plasma generated on NPs and during laser breakdown of the liquid is the source of electrons that may be responsible for water dissociation by direct electron impact. This process starts if the electrons have energy above 16 eV [7].

$e + H_2O \rightarrow$

$H^+ + OH + 2e$ (*Eap* = 16.95 eV)*;*

$H^+ + OH^- + e$ (*Eap* = 16.00 eV)*;*

$H^+ + OH(X2P) + 2e$ (*Eap* = 18.70 eV)*;*

$H_2^+ + O + 2e$ (*Eap* = 20.70 eV)*;*

$HO+ + H + 2e$ (*Eap* = 18.11 eV)*;*

$O^+ + H_2 + 2e$ (*Eap* = 19.00 eV)*;*

$O^+ + 2H + 2e$ (*Eap* = 26.80 eV)

Here the symbol "ap" stands for appearance of corresponding process. One can see that O and H are simultaneously presented in the reaction products. This means that the chemical reaction between them is possible, and formation of water molecules takes place. In turn, this explains why the concentration of molecular hydrogen reaches the saturation level with exposure time (Fig. 3 and 5).

The onset of water breakdown is probably due to the instability of liquid heated by focused laser beam to formation of rapidly expanding bubbles. Let us consider the appearance of small gas bubble in the beam waist in slightly absorbing liquid. This bubble may appear due to dissolved air in the liquid in the beginning of laser exposure or due to dissolved $H_2$ at the late stages of exposure. The absorption could be due to NPs in the liquid, to intrinsic absorption bands of the liquid at laser wavelength, or uncontrollable impurities. The scanning velocity of the laser beam of 1000 mm/s and radius of the circular trajectory of 20 mm corresponds to 60 ms of the dwell time of the laser beam in one particular point of the liquid. The refractive index inside the bubble is lower than in liquid, so some part of the laser beam cannot enter into the bubble. The entered part of the laser beam hits its opposite side and evaporates the liquid inside the bubble. Therefore, the beam radius R will increase with time. The fraction of laser radiation that enters inside the bubble depends on its radius R. This is illustrated in Fig.12 . Some part of the laser beam, which is supposed to have plane wavefront in the beam waist undergoes total reflection. This correspond to the incidence angle $\alpha$, which is determined by the condition: $\sin \alpha = 1/n$, where n stands for the refractive index of the liquid, while the refractive index inside the bubble is equal to 1. The radius h of this "window" for the laser beam can be found as $h = R \sin \alpha$. The area of this circle $S_e = \pi h^2 = \pi R^2/n$. In quasistatic approximation (continuous laser radiation instead of pulsed one) the rate of liquid evaporation into the bubble should be proportional to $S_e$ and local value of average power of laser beam. Therefore, the radius increases with time. The larger is R, the higher is absorbed laser power and, in turn, the rate of evaporation of the liquid inside the bubble. In other words, the system demonstrates positive feedback. The initial radius $R_0$ from which the growth starts should by above the laser wavelength. It may appear from air dissolved in the liquid or from hydrogen at late stages of laser exposure. Otherwise, this initial bubble may originate from a vapor shell that surrounds absorbing impurity, dust or NPs. Probably, the laser breakdown of water starts from the breakdown of the water vapors inside rapidly expanding bubble.

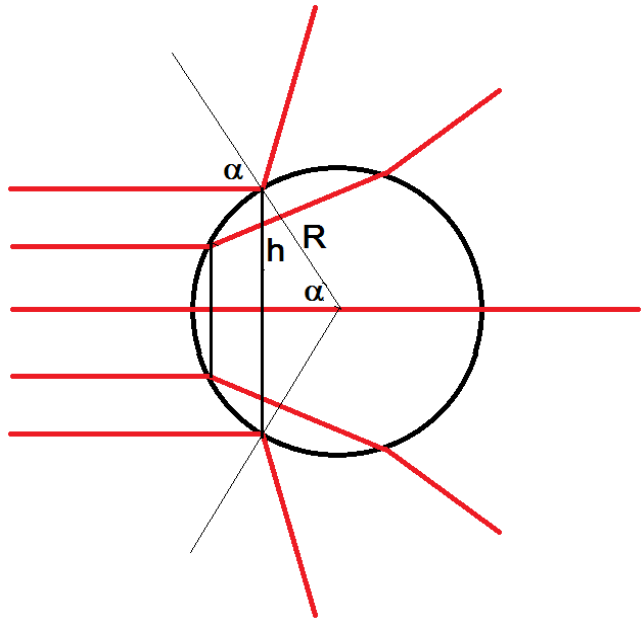

Fig. 12. Bubble of radius R in the laser beam waist. h is the radius of the fraction of laser beam that enters the bubble, other parts of laser radiation is reflected at the water-bubble interface.

Taking into account the obtained results it can be inferred that $H_2$ emission occurs in any experiment on laser ablation in liquids including the ablation of a solid target in the liquid. Preliminary tests on the ablation of a steel target in water confirm this hypothesis. In this point of view the results of previous work on hydrogen emission under laser exposure of carbon particles in water [5] could be re-considered. Indeed, the laser fluence used in this work was of order of 100 mJ/cm$^2$ is not sufficient for water breakdown. On the other hand, this fluence is sufficient for formation of plasma plume on carbon target, and therefore, for dissociation of surrounding water molecules.

Conclusion

Thus, hydrogen emission has been experimentally demonstrated under laser-induced breakdown of colloidal solutions of Au NPs in water. The emission of $H_2$ is clearly related to the breakdown of the liquid and is negligible if plasma is not ignited. The $H_2$ emission is observed in the purest available samples of waters that contain no controllable solid impurities. $H_2$ emission is tentatively assigned to $H_2O$ molecules dissociation by electron impact from plasma of laser breakdown.

Acknowledgments

The authors gratefully acknowledge the support of the RF President's Grants Council for Support of Leading Scientific Schools (Grant # NSh-4484.2014.2), Russian Foundation for Basic Researches, Grants 15-02-04510_a, 15-32-20926_mol_a_ved, 16-02-01054_a, and RF President's Grant MK-4194.2015.2.